# Soil moisture estimation of bare and vegetation-covered areas using a P/L/C-band SAR

Gian Oré, Jhonnatan Yepes, Juliana A. Góes, Luciano P. Oliveira, Bárbara Teruel and Hugo E. Hernandez-Figueroa,
*Senior Member, IEEE*

*Abstract*— **The paper introduces a novel approach for estimating soil moisture in vegetated surfaces, specifically focusing on sugarcane crops throughout various growth stages in agriculture applications. While existing models typically address bare soil scenarios, this model utilizes data from P-, L-, and C-band Synthetic Aperture Radar (SAR) to estimate soil moisture. The semi-empirical Dubois model forms the basis of the proposed model, adapted to accommodate multiband operation and crop height variations. Synthetic datasets are generated using the adjusted model for training two neural networks incorporated into the overall model. Additionally, a linear expression for estimating crop height is integrated into the model. The model is validated in an Experimental Site at the School of Agricultural Engineering, UNICAMP, and an independent area at the Sugarcane Technology Center in Piracicaba, Brazil. The model utilizes a multiband drone-borne SAR system with a 3-meter image resolution and radiometric accuracy of 0.5 dB. The results indicate that the model can estimate soil moisture with root-mean-square errors of 0.05 $cm^3.cm^{-3}$ (5 vol. %) across crop heights ranging from zero to 2.5 meters.**

*Index Terms*—**Drone-borne radar, neural networks, precision agriculture, sugarcane, soil humidity, synthetic aperture radar.**

## I. INTRODUCTION

Soil moisture content is critical to several processes, including agricultural, agronomic, geological, ecological, biological, and hydrological [1]. Spatial and temporal soil moisture knowledge is paramount for irrigation and crop development [2], [3]. Such information is vital for sugarcane crop management, which is complex due to the rising demand for derivatives and the continual expansion of cultivation areas [4]. This is especially true in Brazil, one of the world's top sugar and ethanol producers [5].

Many studies strive to overcome such limitations by estimating soil moisture from synthetic aperture radar (SAR) data, exploiting SAR's ability to map large areas with high temporal and spatial resolution [7], [8].

Empirical and semi-empirical models based on SAR images are commonly used for soil moisture estimation in bare soil scenarios [7]. Empirical models based on a single polarization and angle of incidence are more effective with short wavelengths and low incidence angles to minimize the effects of ground roughness on radar reflectivity during data acquisition [9]. The semi-empirical model relates the reflectivity with the sensor specifications (frequency, polarization, incidence angle) and the surface parameters (soil roughness and moisture). Several semi-empirical models have been published, and among the most popular are those developed by Oh *et al.* [10], Dubois et al. [11], Baghdadi *et al.* [12], Mirsoleimani *et al.* [13], Gorrab *et al.* [14], and Ponnurangam *et al.* [15]. It introduces fewer restrictions than the empirical model and offers more degrees of freedom to model the radar reflectivity. Some methods use polarimetric data to estimate soil moisture, while others use neural networks. Some approaches consider different scenarios with vegetation. However, these models have never been validated for the P-band or other low UHF frequencies.

In most agriculture applications, vegetated surfaces are the norm. Radar reflectivity in such conditions comprises signals bounced back from the soil, vegetation, and their interaction. The radar wavelength determines the signal attenuation, with lower attenuation occurring for higher wavelengths, such as the C band. Most works that estimate soil moisture from radar data use the C band, as most of these sensors are open-access [6, 12, 13, 15]. Also, attenuation is caused by the vegetation layers, resulting in a lower soil contribution when vegetation is denser. Such an attenuation increases as the crop grows and becomes more intense for high-incidence angles [16], [17]. According to some studies, L-band SAR can be used to estimate soil moisture in vegetated surfaces. The leaf area index and normalized difference vegetation index are also considered to account for the vegetation effect [18, 19]. The authors of these studies report root-mean-square errors (RMSE) of 0.08 and 0.046 $cm^3.cm^{-3}$ (cubic centimeters of water per cubic centimeter of soil). Park *et al.* [20] used a time-series method to estimate soil moisture in vegetation scenarios. They assumed that soil roughness and vegetation properties remain consistent over two consecutive measurements. Another work based on time series to estimate moisture was proposed by Burgi *et al.* [21], which uses InSAR information from C band radar. Lal *et al.* [22]

Gian Oré and Hugo E. Hernandez-Figueroa are with the School of Electrical and Computer Engineering, University of Campinas–UNICAMP, Campinas 13083-852, Brazil (emails: g228005@dac.unicamp.br; hugo@unicamp.br).

Jhonnatan Yepes and Bárbara Teruel are with the School of Agricultural Engineering, University of Campinas–UNICAMP, Campinas, 13083-875,

Brazil (email: j261219@dac.unicamp.br; barbarat@unicamp.br).

Juliana A. Góes is with the Campinas Agronomic Institute, Campinas, 13075-630, Brazil (email: jugoes.eng@gmail.com).

Luciano P. Oliveira is with the Directed Energy Research Centre, Technology Innovation Institute, Abu Dhabi P. O. Box 9639, United Arab Emirates (email: Luciano.Prado@tii.ae).

Color versions of one or more of the figures in this article are available online at http://ieeexplore.ieee.org.

proposed a multi-scale algorithm for soil moisture from UAVSAR L-band data for various vegetated scenarios. Such an algorithm doesn't require time series observations or ancillary data. A lower frequency SAR system like P-band SAR in well-developed agricultural fields can monitor soil moisture with added degrees of freedom. The longer wavelengths of P-band SAR can penetrate through the plant's canopy, stems, and leaves, providing more accurate soil information as the crops grow [23], [24]. Scenarios involving vegetation have been modeled [15], but these models have never been validated for low UHF bands like P-band.

The paper introduces a novel method to estimate soil moisture in sugarcane plantations at any growth stage using a three-band SAR system. In addition to the L- and C-band SAR data used in previous works [11], [12], the proposed model incorporates P-band data from the multiband drone-borne SAR system used in flight surveys. Dubois' semi-empirical is extended with a term relating total reflectivity to signal wavelength and crop height to enhance soil moisture estimation accuracy in vegetated surface scenarios. The adjusted Dubois model generated synthetic data sets to train neural networks. These networks provide a single estimation of soil moisture from three frequency bands. The complete estimation model also includes a linear model that determines crop height.

The paper is organized as follows. Section II describes the study sites and classification of the SAR datasets. Section III details the adjustments made to the original Dubois model to make it applicable in a vegetated surface scenario. The proposed model for estimating soil moisture is broken down in Section IV. Section V presents the results of the complete estimation model at the Experimental and Testing Sites. Finally, Section VI presents the Conclusions.

## II. STUDY SETTINGS

### A. Experimental Site

The proposed model for soil moisture estimation was fitted using data from bare soil and vegetated surface scenarios at a 10 x 40 m² Experimental Site, shown in Fig. 1. The site is located at the School of Agricultural Engineering of the University of Campinas (FEAGRI/UNICAMP), Campinas, São Paulo (22°49' 12"S, 47° 3' 42"W). According to the USDA soil classification system, the soil is composed of clayey oxisol soil type, with 25% sand and 57% clay [25]. A micro-sprinkler system was installed in June 2019 to induce variations in soil moisture. In July 2019, 340 sugarcane seedlings (type IACSP97-4039) were planted in seven rows, spaced 1.5 and 0.75 meters apart.

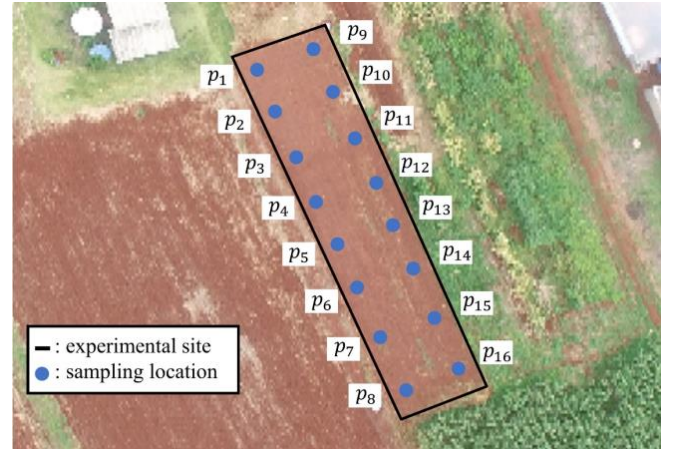

**Fig. 1.** Optical image of the Experimental Site.

The Experimental Site surveys consisted of three activities, summarized in Table I: an SAR survey, soil sampling, and sugarcane biometric measurement (height measurements). The soil samples were collected from 0 to 20 cm depth at the points shown in Fig. 1. On the same day as collection, the laboratory assessed the gravimetric moisture content.

TABLE I
DATA SURVEYS

| Sets | Survey dates | Mean soil moisture (cm³.cm⁻³) (measured) | Mean crop height (m) (measured) |
|---|---|---|---|
| Set A | 02 July 2019 | 0.3919 | 0.0 |
| | 17 July 2019 | 0.3452 | 0.0 |
| | 24 October 2019 | 0.2911 | 0.11 |
| | 18 November 2019 | 0.3466 | 0.21 |
| Set B | 19 December 2019 | 0.3490 | 0.56 |
| | 04 March 2020 | 0.3278 | 1.86 |
| | 10 April 2020 | 0.3033 | 2.24 |
| | 12 May 2020 | 0.2019 | 2.29 |

The data was divided into two sets based on the sugarcane height: Set A below 0.5 m and Set B above 0.5 m. The data from the sampling location shown in Fig. 1 was also split into tuning and validation datasets, from $p_1$ to $p_8$, used for model tuning, whereas data from $p_9$ for $p_{16}$ for validation.

### B. Testing Site

The fitted model for soil moisture estimation was tested on a 250 x 80 m² site shown in Fig. 2. This Testing Site is owned by the Sugarcane Technology Center (CTC) in Piracicaba, Sao Paulo. The site holds various types of sugarcane plants used for research. On 25 May 2022, SAR data was collected when the sugarcane plants had an average height of 2 meters. Soil samples were taken from 10 locations on the same day to determine gravimetric soil moisture at 0-20 cm depth using the same procedure as for the Experimental Site.

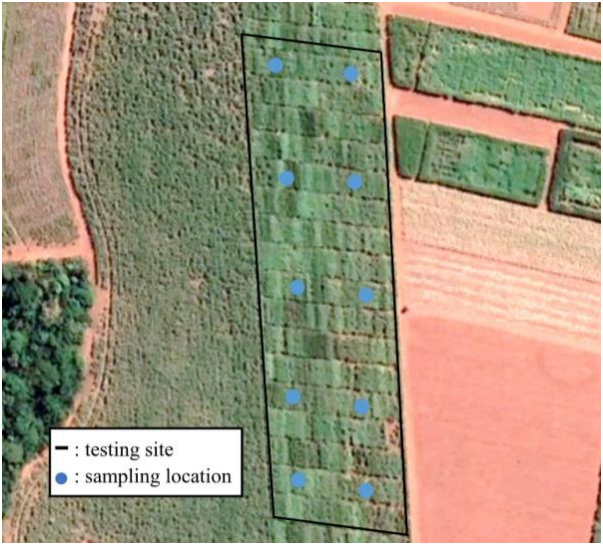

**Fig. 2.** Optical image of the Testing Site.

*C. SAR Imaging*

The SAR surveys were carried out by a drone-borne SAR system operating in the P-, L-, and C-bands, with the acquisition parameters presented in Table II and described in [26]. All Synthetic Aperture Radar (SAR) surveys were conducted using linear flights. The drone's average flight height was 120 meters, with a swath width of 280 meters. The near range was 130 meters at a 20-degree incidence angle, while the far range was 350 meters at a 70-degree incidence angle. The Experimental Site was illuminated between a range of 230 m at a 59-degree incidence angle and 266 m at a 63-degree incidence angle.

It is worth noting that drone-based and airborne Synthetic Aperture Radar (SAR) typically use an incidence angle of around 60 degrees, which is the main focus of this publication. However, drone-borne surveys can also simulate satellite-borne SAR scenarios with an incidence angle of 30 degrees, which is common for satellite-based applications.

The acquired data were processed with a time-domain back-projection algorithm [27]. After processing, the speckle noise was reduced by applying a moving average filter of 1.5 x 1.5 m². Furthermore, corner reflectors with known theoretical radar cross-sections were placed in the Experimental Site for absolute radiometric calibration, which was conducted using the integral method [28] and achieved a 0.5 dB accuracy. Finally, the average reflectivity value of each sampling location was computed using 3 x 3 m² windows centered on the SAR image's central pixel.

TABLE II
DRONE BORNE SAR SYSTEM PARAMETERS

| Radar Parameters | P band | L band | C band |
|---|---|---|---|
| Polarization | HH | HH | VV |
| Wavelength | 70.5 cm | 22.8 cm | 5.6 cm |
| Bandwidth | 50 MHz | 150 MHz | 200 MHz |
| Azimuth beamwidth | 55.9° | 58.5° | 32.5° |
| Elevation beamwidth | 69.3° | 79.8° | 51.3° |
| Azimuth resolution | 0.3 m | 0.1 m | 0.05m |
| Range resolution | 4 m | 1.2 m | 1 m |

*D. Soil Moisture Measurement*

Volumetric soil moisture (K) is a parameter obtained from all indirect methods used to infer soil property. It expresses the relationship between the volume of water present in the soil and the total soil volume measured in cm³.cm⁻³. Volumetric soil moisture can be defined as a function of gravimetric soil moisture (W), bulk soil density ($\rho_a$), and soil water density ($\rho_w$) [29] as:

$$K = W * \frac{\rho_a}{\rho_w}, \tag{1}$$

The gravimetric soil moisture was determined by extracting soil samples from a depth of 0 to 20 cm using a manual drill and then following the gravimetric method with oven drying [29,30]: weighing each moist sample, oven drying them at a temperature of $100 \pm 5°C$ over 24 to 48 hours, and weighing each sample again, once they are dried. The bulk soil density was determined by collecting a sample from the center of the Experimental and Testing Sites and assuming uniformity across each area.

*E. Other Measurements*

Local soil roughness, $h_{rms}$, was estimated to be 2.21cm in the experiment and assumed to be uniform across all sampling locations due to the Experimental Site's homogeneity. The cane seedlings were planted, and biometric measurements were taken to calculate biomass and crop height. The methodology for biometric measurements is detailed in [26].

## III. ADJUSTED DUBOIS MODEL

This section presents a modified version of the Dubois model to suit a vegetated surface scenario. The adjustments consider the varying penetration capabilities of the P-, L-, and C-bands used in the drone-borne SAR system. Throughout the phenological stage of a crop, each band acquires greater importance according to its penetration capacity. In particular, the sugarcane crop is almost transparent to the P-band, a promising feature for estimating soil moisture on vegetated surfaces. However, the P-band was not considered in the original Dubois model. The proposed adjustments also profit from the drone-borne SAR system's high-resolution images.

*A. The Dubois Model*

Dubois *et al.* [11] presented a semi-empirical model to estimate bare soil moisture. The model describes the dependence of the co-polarized reflectivity ($\sigma^0$) on sensor-surface parameters, such as incident angle ($\theta$), soil dielectric constant ($\varepsilon$), and soil roughness ($h_{rms}$), and is mathematically expressed as [11]:

$$\sigma_{HH}^0 = f_H(\theta) \, g(h_\lambda, \theta)^{a_H} \, m_H(\varepsilon, \theta) \, \lambda^{b_H}, \tag{2}$$

$$\sigma_{VV}^0 = f_V(\theta) \, g(h_\lambda, \theta)^{a_V} \, m_V(\varepsilon, \theta) \, \lambda^{b_V}, \tag{3}$$

where $\sigma_{HH}^0$ and $\sigma_{VV}^0$ are, respectively, the direct horizontal and vertical reflectivity, $\lambda$ is the wavelength in centimeters, $h_\lambda = kh_{rms}$ is the normalized soil roughness. Also [11]:

$$g(h_\lambda, \theta) = h_\lambda \sin\theta, \tag{4}$$

$$m_H(\varepsilon, \theta) = 10^{(c_H \varepsilon \tan \theta + d_H)}, \tag{5}$$

$$m_V(\varepsilon, \theta) = 10^{(c_V \varepsilon \tan \theta + d_V)}, \tag{6}$$

$$f_H(\theta) = \cos(\theta)^{1.5}/\sin(\theta)^5, \tag{7}$$

$$f_V(\theta) = \cos(\theta)^3/\sin(\theta)^3. \tag{8}$$

The variables $a_H$, $b_H$, $c_H$, $d_H$, $a_V$, $b_V$, $c_V$, and $d_V$ are fitting constants. The Dubois model is valid for volumetric soil moisture $m_v < 0.35$ cm³.cm⁻³, normalized soil roughness $kh_{rms} < 3$, and incident angle $\theta > 30°$.

The Dubois model was originally designed to process data separately from L-, C-, or X-bands. In contrast, the adjustments proposed in the following sections consider the multiband information offered by the drone-borne SAR system. Consequently, the adjusted Dubois model is only on eq. (2), and applied for the following frequencies and polarizations: C-VV, L-HH, and P-HH.

$$\sigma^0_{(D)} = f(\theta) \, g(h_\lambda, \theta)^a \, m(\varepsilon, \theta) \, \lambda^b, \tag{9}$$

$$f(\theta) = \cos(\theta)^{1.5}/\sin(\theta)^5, \tag{10}$$

$$m(\varepsilon, \theta) = 10^{(c\varepsilon \tan \theta + d)}, \tag{11}$$

### B. Crop Height Adjustment

As the Dubois model cannot be applied when the soil is covered with vegetation, the first adjustment considers the effect of vegetation on total reflectivity. This is achieved by adding to Eq. (09) the term $n_H(h)$, representing the reflectivity contribution due to the crop height, as follows:

$$\sigma^0_{(12)} = [f(\theta) \, g(h_\lambda, \theta)^a \, m(\varepsilon, \theta) \, \lambda^b] \, n_H(h). \tag{12}$$

Fig. 3 compares the measured crop height against the reflectivity from all datasets from the Experimental Site.

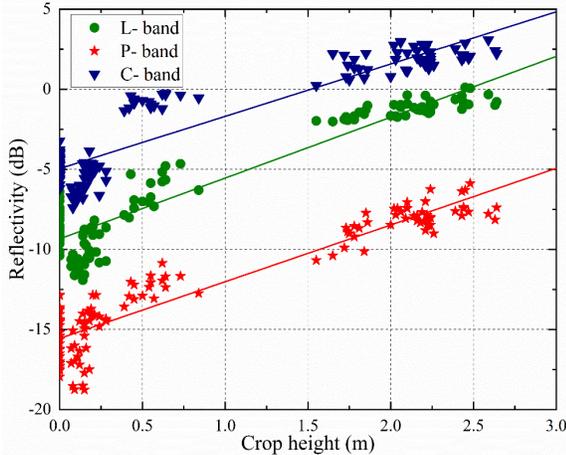

**Fig. 3.** Measured crop height and reflectivity values for P-, L- and C- bands using sets A and B.

Note that the reflectivity of the C band increases rapidly with the height of vegetation, and from 0.5m onwards, the reflectivity increases slowly. This effect is minor for L-band data and is not noticeable for P-band data. Although a linear model does not coincide correctly with the C band, other higher-degree models were not considered because it is not used when the vegetation is dense due to its saturation with dense vegetation. The L and P bands play an important role as they do not saturate up to 2.5 m. The relationship between crop height and reflectivity (in decibels) can be expressed by a linear model. However, since (12) is not in decibels, the relationship must be adjusted to:

$$n_H(h) = 10^{(a_0 h + b_0)}, \tag{13}$$

where $h$ is the crop height, and $a_0$ and $b_0$ are first-order model coefficients. The term $n_H(h)$ represents vegetation's height contribution to reflectivity. Thus, Equation (12) describes the total reflectivity as a function of soil moisture, soil roughness, angle of incidence, and vegetation height.

### C. Wavelength Adjustment

The original Dubois model needs a specific set of fitting constants for each operating band and polarization. However, the adjusted model uses only one set of constants, which means that adding a term, $n_W(\lambda)$, to improve control over the wavelength's effect is necessary. Thus, the adjusted Dubois model can be expressed in its final form as

$$\sigma^0 = [f_H(\theta) \, g(h_\lambda, \theta)^a \, m_H(\varepsilon, \theta) \, \lambda^b] \, n_H(h) n_w(\lambda) \tag{14}$$

where $n_w(\lambda)$ is the wavelength effect parameter. This term allows for greater control over wavelengths' contributions to total reflectivity. Although the original Dubois model includes a term for the wavelength effect, it was barely tested for the L, C, and X bands.

To determine $n_w(\lambda)$, equation (12) was empirically adjusted using the minimum mean-square-error (MMSE) criterion for all datasets from the Experimental Site. The difference between the measured reflectivity and the estimated reflectivity from (12) can be calculated as:

$$\sigma^0_{(dif)} = \sigma^0_{(S)} - \sigma^0_{(12)}, \tag{15}$$

where $\sigma^0_{(S)}$ represents the reflectivity measured from SAR images and $\sigma^0_{(12)}$ is the reflectivity estimated from (12). A boxplot of Equation (15) results as a function of wavelength is shown in Fig. 4. The measured reflectivity was obtained from SAR images using all available data.

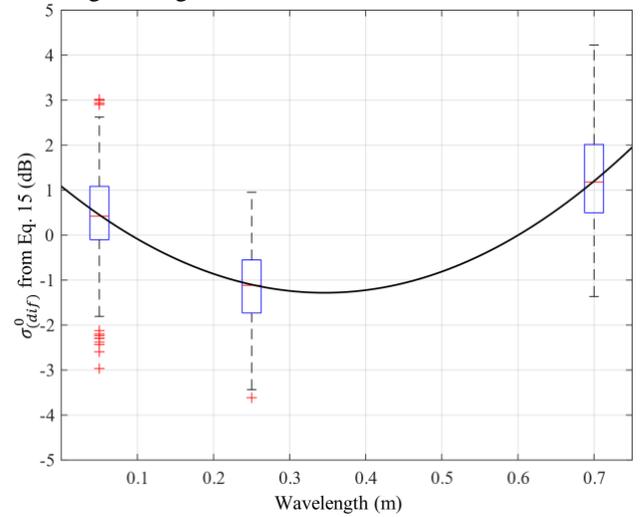

**Fig. 4.** Boxplots of reflectivity difference as a function of the wavelength.

The difference between measured and estimated reflectivity (in decibels) presents a quadratic trend, as shown by the black line in Fig. 4. This effect can be modeled as:

$$n_w(\lambda) = 10^{(c_0 \lambda^2 + d_0 \lambda + e_0)}, \tag{16}$$

where $c_0$, $d_0$ and $e_0$ are coefficients of a second-order model. Since (13) already contains a zero-order coefficient ($b_0$), $e_0$ can be included in $b_0$.

Finally, the fitting constants in (13), (14), and (16) were calculated based on the MMSE criterion using data from P, L, and C-band tuning datasets. The resulting constants are $a = 1.4$, $b = 0.47$, $c = 0.014$, $d = -0.72$, $a_0 = 0.42$, $b_0 = 0.17$, $c_0 = -2.4$ and $d_0 = 1.76$.

## IV. STRUCTURE OF THE SOIL MOISTURE ESTIMATION MODEL

The complete proposed model for soil moisture estimation includes different components, allowing it to work for both bare soil and vegetated surfaces. The block diagram in Fig 5 illustrates how the soil moisture ($m_v$) is estimated from the input radar reflectivity ($\sigma_P^0$, $\sigma_L^0$, $\sigma_C^0$) and inclination angle ($\theta$). First, the master algorithm estimates the crop height (H-LM) to determine whether to use a neural network for calculating the final soil moisture for bare soil (B-NN) or vegetated surfaces (V-NN). B-NN and V-NN are Multilayer Perceptron neural networks [13], [18] providing one output (soil moisture) from the multiband input data. Essentially, they invert the adjusted Dubois model.

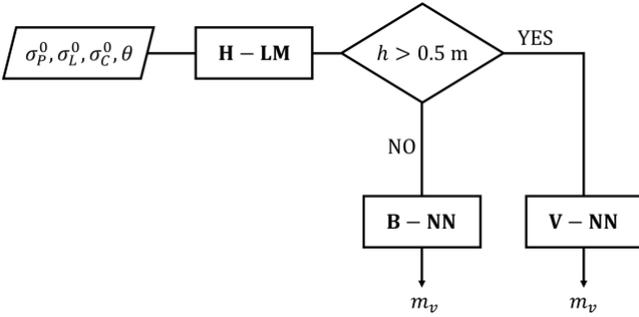

**Fig. 5.** Block diagram of the complete model for soil moisture estimation.

### A. Crop Height Estimation (H-LM)

Fig. 3 shows a clear linear relationship between crop height and reflectivity data. Therefore, a linear model (H-LM) was selected to estimate crop height. This estimation model only considers P and L-band data, as the C-band has saturation problems when dealing with dense vegetation. The equation of the linear model H-LM is expressed as

$$h = 3.119 + 0.1372\sigma_L + 0.1117\sigma_P \qquad (17)$$

where $\sigma_L$ and $\sigma_P$, in decibels, are HH reflectivity. Equation (17) was fitted using the tuning datasets from sets A and B.

Fig. 6 compares the crop height measured from all data to the estimated crop height using the H-LM. The Root-Mean-Square Error (RMSE) of the tuning and validation datasets is 0.31 and 0.25 meters, respectively. The estimated crop height is utilized as a control input and as input data for the B-NN and V-NN, as shown in Fig. 5.

### B. Moisture Estimation for Bare Soil (B-NN)

The bare soil scenario's neural network (B-NN) has an input layer with six neurons, two hidden layers of twenty neurons each, and an output layer with a single neuron. The neurons in the hidden layers use the hyperbolic tangent-sigmoid activation function, and the output layer uses a linear activation function.

The Levenberg-Marquardt algorithm [31] was used to train the neural network to minimize the error between the predicted outputs provided by the B-NN and the reference output.

The B-NN inputs are the incidence angle, ground roughness, crop height, P-, L- and C-band reflectivity. The soil moisture is the output. The B-NN was trained using a synthetic dataset generated using the adjusted Dubois model to produce reflectivity values for the P-, L-, and C-bands. The synthetic dataset was created using incidence angles ranging from 60 to 65 degrees, soil moisture ranging from 0.05 to 0.45 cm$^3$.cm$^{-3}$, soil roughness ranging from 1.5 to 3.5 cm, and crop height ranging from zero to 0.5 m.

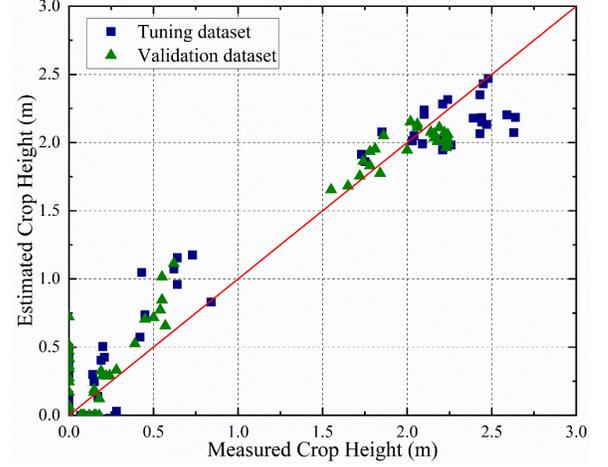

**Fig. 6.** Crop height: measured vs. estimated with the H-LM (datasets A and B).

The B-NN was validated using biometric measurements, even though the complete method estimates crop height. Fig. 7 displays the soil moisture that was estimated by the B-NN using Set A data (Table 1) and compares it to the actual measured values. The RMSE for the tuning and validation datasets are 0.03 and 0.0315 cm$^3$.cm$^{-3}$, respectively. The RMSE values obtained by the B-NN in the validation dataset are consistent and fall within the same range as those achieved by other authors [13]–[15], although they used C and X frequency bands with incidence angles of 30 to 40 degrees in their works.

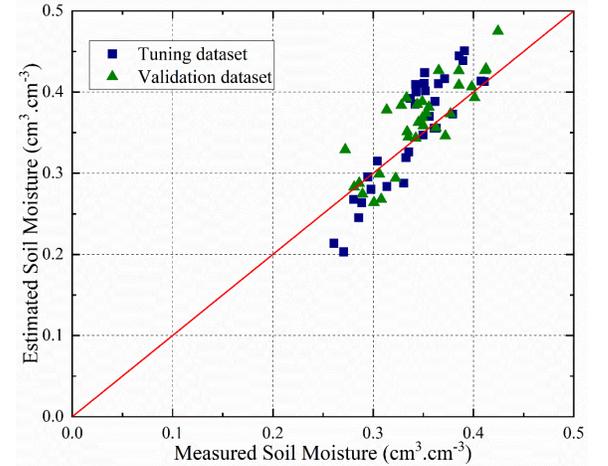

**Fig. 7.** Soil moisture: measured vs. estimated with the B-NN (set A data).

### C. Moisture Estimation for Vegetated Surfaces (V-NN)

When dealing with vegetated surfaces, the reflectivity increases with the height of the crop. Furthermore, the impact of vegetation is more evident for shorter wavelengths. For this reason, the C-band reflectivity was not used here due to the saturation issues typical in scenarios with dense vegetation [16], [17].

The V-NN estimates soil moisture under vegetation using incidence angle, ground roughness, crop height, and P- & L-band reflectivity as inputs and soil moisture as the single output. The V-NN has an input layer of five neurons, two hidden layers of twenty neurons each, and an output layer of one neuron. Other characteristics of the V-NN are similar to the B-NN. The V-NN was trained with a synthetic dataset based on the adjusted Dubois model to generate P- and L-band reflectivities. The V-NN synthetic dataset used the same incidence angle, soil moisture, and soil roughness values as the B-NN. However, the crop height varied from 0.5 to 3.0 m.

Upon validation using biometric measurements, the V-NN was compared with the measured soil moisture to obtain estimated values. Fig. 8 compares the measured soil moisture with estimated values derived from the V-NN using Set B data presented in Table 1. The RMSE for the tuning dataset is 0.0473 cm³.cm⁻³, while for the validation dataset, it is 0.0514 cm³.cm⁻³. These values are higher than those obtained for the bare soil scenario, which can be attributed to the presence of vegetation. In a vegetated scenario, the L and P bands exhibit slightly different behavior, with the L band having lower penetration. Additionally, both bands may not be sensitive to the same soil layer.

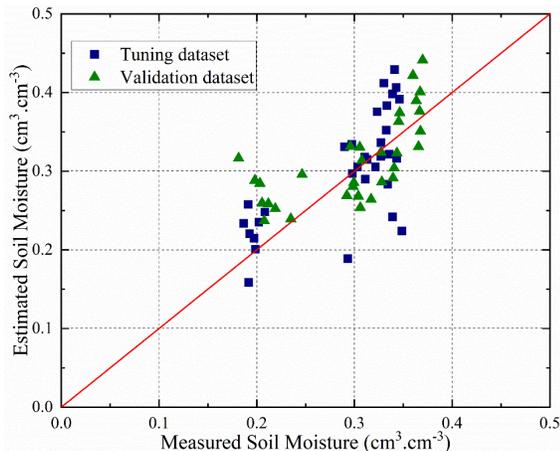

**Fig. 8.** Soil moisture: measured vs. estimated with the V-NN (set B data).

## V. COMPLETE MODEL FOR SOIL MOISTURE ESTIMATION

This section presents the results of the complete model shown in Fig. 5, which estimates soil moisture. The model was assessed using data from Experimental and Testing sites.

### A. Experimental site

Figure 9 compares the soil moisture measured at the Experimental Site to the values estimated by the complete soil moisture model. The RMSE is approximately 0.05 cm³.cm⁻³: 0.0457 and 0.0501 cm³.cm⁻³ for the tuning and validation datasets. The RMSE is 0.0314 and 0.0518 cm³.cm⁻³ for Sets A and B, respectively.

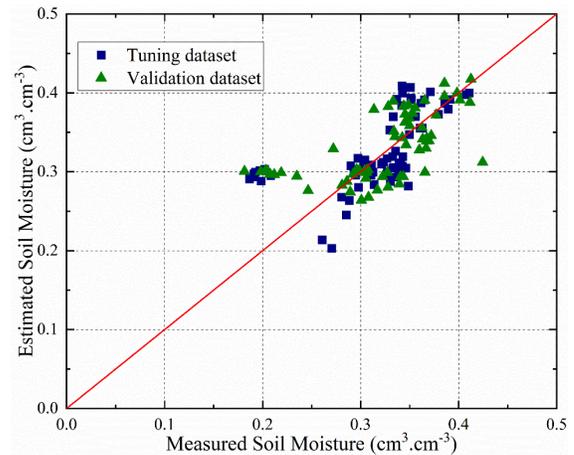

**Fig. 9.** Soil moisture at the experimental site: measured vs. estimated using the complete model (datasets A and B).

The complete moisture estimation model performs better in bare soil, which agrees with the results presented in Fig. 7 and 8, where the V-NN produces more scattered values than the B-NN.

Fig. 10 presents soil moisture maps for the experimental site with (a) bare soil and (b) a 2.24-m crop height. Both maps exhibit considerable heterogeneity, which is to be expected as the micro-sprinkler system was configured to induce soil moisture variations. The estimated soil moisture average is slightly lower in the northern area compared to the southern area. The measurements are 0.30 cm³.cm⁻³ for bare soil, as shown in Fig. 10a, and 0.38 cm³.cm⁻³ for the southern area. Similarly, for sugarcane coverage, the average soil moisture level is 0.24 cm³.cm⁻³ for the northern area and 0.32 cm³.cm⁻³ for the southern area, as shown in Fig. 10b. These results indicate that the soil moisture level is reduced on average when there is sugarcane coverage.

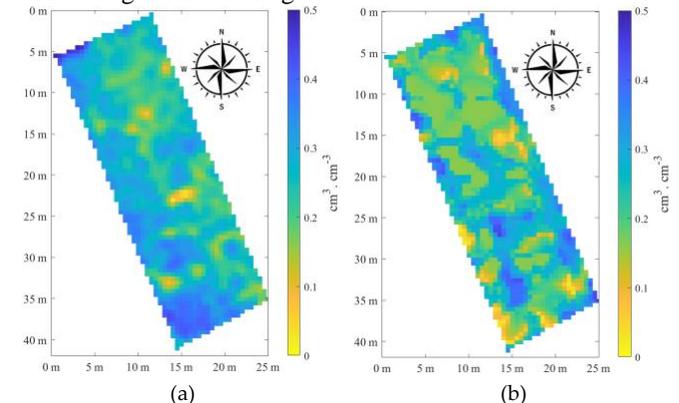

**Fig. 10.** Maps of estimated soil moisture at the Experimental Site: (a) bare soil on July 17, 2019, and (b) vegetated surface on April 10, 2020.

The maps show a wide range of soil moisture values, which can be attributed to the SAR images' high resolution and the Experimental Site's heterogeneity. Some regions exhibit very low humidity values due to speckle, non-uniform soil roughness, or inaccurate vegetation height estimation, as seen in Fig. 10b.

### B. Testing site

Additionally, the complete model was evaluated at the Testing Site. In this case, 10 soil samples were taken to determine moisture, and only the vegetated surface scenario was considered. Fig. 11 compares measured and estimated soil moisture with an RMSE of 0.0488 cm³.cm⁻³. This level of accuracy is similar to the results obtained from the Experimental Site. Fig. 12 displays the estimated moisture map of the Testing Site.

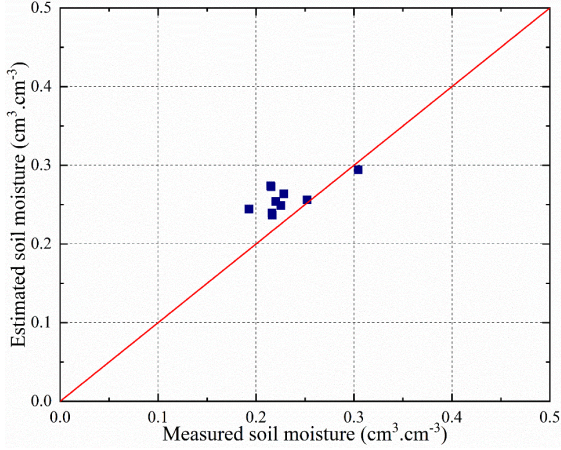

**Fig. 11.** Soil moisture at the Testing Site: measured vs. estimated using the complete model.

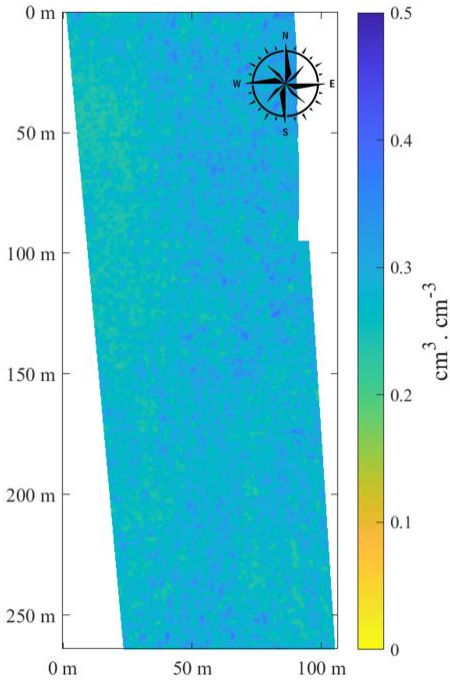

**Fig. 12.** Estimated moisture map of the Testing Site.

Moreover, the completed model was successfully validated with the same equations and parameters in a pulp industry farm with 4 soil samples, a 30 % to 45% moisture range, and a resulting RMSE of 8.13 %. Grouping this data set with one of the Testing Sites at CTC, we obtained an RMSE of 5.77 % based on 14 samples. This algorithm has been operationally used by drone-borne radar survey enterprises [32] and presents a very stable behavior with an RMSE of 5 %, always using the same model structure and parameters presented here. The same bands are used for biomass estimation, digital surface, and terrain model estimation [33] so that the same survey can evaluate a crop inventory and the soil moisture map.

Out of the three bands available in the SAR system, band C is ineffective in estimating soil moisture when significant vegetation exists. Therefore, the vegetation height is used as input information to determine the scenarios where considering the C band is useful. This helps to decide whether to use the output of B-NN or V-NN. It is worth noting that when dealing with a flat, vegetation-free surface with gravimetric soil moisture below 15%, the C band provides an important and accurate contribution. In contrast, L and P bands present low reflectivity and are not suitable to obtain gravimetric soil moisture RMSE of about 5%. L-band penetrates up to 20 cm and plays an important role in gravimetric soil moisture range of 10-30%, even in the presence of vegetation. The P-band is particularly useful in areas with dense vegetation, especially when the soil moisture is about 20% based on gravity measurements.

## VI. CONCLUSIONS

The paper introduces an innovative method for estimating soil moisture in sugarcane plantations, utilizing three-band Synthetic Aperture Radar (SAR) data and neural network algorithms. The approach is versatile, applicable to both bare and vegetated surface scenarios across various growth stages. Crop height estimation is facilitated by a linear model (H-LM) within the comprehensive model, guiding the selection of soil moisture from either the bare soil neural network (B-NN) or the vegetated surface neural network (V-NN). Neural networks play a crucial role in inverting the adjusted Dubois model, producing a unified outcome from diverse reflectivity inputs of different frequency bands, aiming for optimal results.

The complete model demonstrates a satisfactory root-mean-square error (RMSE) of 5 vol. % or 0.05 cm³.cm⁻³. Focused on B-NN and V-NN, the estimations yield RMSE values of 0.0315 cm³.cm⁻³ and 0.0514 cm³.cm⁻³, respectively, at the Experimental Site.

The multiband approach employed in V-NN effectively mitigates the impact of sugarcane cultivation on soil moisture estimates, maintaining accuracy for crop heights under 2.25 meters. The synergistic use of P-band and L-band proves essential for V-NN, addressing saturation issues up to 2.25 meters and complementing each other. The roles of L-band for lower crop heights and P-band for higher crop heights are emphasized, while the exclusion of C-band minimally affects the model's performance and is exclusively utilized in B-NN.

The multiband drone-borne SAR system provides images with resolutions ranging from one to three meters and 0.5 dB radiometric accuracy, meeting the requirements for the Experimental and Testing Sites with limited variations in soil moisture (0.17 to 0.37 cm³.cm⁻³). Importantly, the proposed model extends its applicability to SAR images with planimetric resolutions spanning from three to hundreds of meters.

Uniform linear flight paths with single polarizations for each band were used in all surveys. Future work considerations include exploring different polarizations, more complex flight paths (e.g., circular or helical trajectories), and the potential integration of tomographic mapping.

## ACKNOWLEDGMENT

This work was partially supported by the São Paulo Research Foundation (FAPESP) under the projects 2021/06506–0 (StReAM), 2021/11380–5 (CPTEn), and 2021/00199–8 (SMARTNESS); the Brazilian Agency CNPq, under the project 312714/2019–0 (HEHF's research productivity grant).

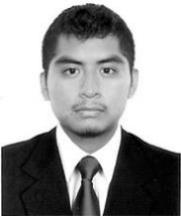

**Gian Oré**, BSc degree in Electronic Engineering from Universidad Peruana de Ciencias Aplicadas (UPC), Lima, Perú, in 2016, and the MSc degree in Electrical Engineering from the University of Campinas (UNICAMP), Campinas, Sao Paulo, Brazil in 2021. He is currently pursuing a PhD degree in Electrical Engineering at UNICAMP. His research interests include remote sensing, interferometry SAR, signal processing, and deep Learning for SAR images.

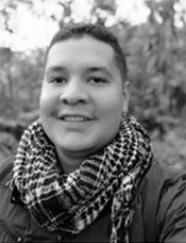

**Jhonnatan Yepes,** MSc in Agricultural Engineering from UNICAMP and Agricultural Engineer from the National University of Colombia (UNAL). Currently, he is a PhD candidate in Agricultural Engineering at UNICAMP. His research focuses on the application of machine learning techniques and data science to optimize agricultural processes and the development of components used in irrigation systems for a efficient application of water resources.

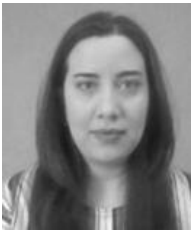

**Juliana A. Góes**, received a PhD degree in Electrical Engineering at the University of Campinas in 2022.

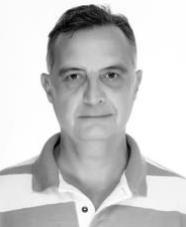

**Luciano P. Oliveira**, PhD in Electrical Engineering from the University of Campinas (UNICAMP), Campinas, Sao Paulo, Brazil, in 2013. Currently, he is a Lead Researcher at the Direct Energy Research Center at the Technology Innovation Institute (TII) in Abu Dhabi, UAE, and a Collaborating Researcher at the School of Electrical and Computer Engineering at UNICAMP.

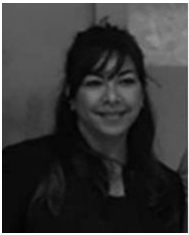

**Bárbara Teruel**, PhD, Mechanical Engineering, Associate Professor in Agricultural Engineering College, at the Laboratory of Energy and Digital Agriculture, in Universidade Estadual de Campinas, Brazil.

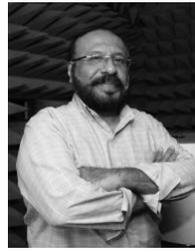

**Hugo E. Hernandez-Figueroa**, PhD in physics from the Imperial College London, in 1994. Full Professor at the University of Campinas (UNICAMP), School of Electrical and Computer Engineering (FEEC), Campinas, Sao Paulo, Brazil, since 2005. Dean of FEEC (2023-2027). Fellow of OPTICA (2011) and a recipient of the IEEE Third Millennium Medal in 2000. Advisory Committee Member in Engineering at the Sao Paulo State Science Foundation (FAPESP) since 2014. His research interests include integrated photonics, radars, photonics biosensors, and metamaterials for communications and biotechnology.